# Credit Risk Assessment Model for UAE's Commercial Banks: A Machine Learning Approach

**Aditya Saxena***[0000-0003-2204-9035], *Birla Institute of Technology and Science, Pilani*
E-mail: f20190089@dubai.bits-pilani.ac.in
**Dr Parizad Dungore**[0000-0002-1943-8426], *Birla Institute of Technology and Science, Pilani*

## Abstract

Credit ratings are becoming one of the primary references for financial institutions of the country to assess credit risk in order to accurately predict the likelihood of business failure of an individual or an enterprise. Financial institutions, therefore, depend on credit rating tools and services to help them predict the ability of creditors to meet financial persuasions. Conventional credit rating is broadly categorized into two classes namely: good credit and bad credit. This approach lacks adequate precision to perform credit risk analysis in practice. Related studies have shown that data-driven machine learning algorithms outperform many conventional statistical approaches in solving this type of problem, both in terms of accuracy and efficiency. The purpose of this paper is to construct and validate a credit risk assessment model using Linear Discriminant Analysis as a dimensionality reduction technique to discriminate good creditors from bad ones and identify the best classifier for credit assessment of commercial banks based on real-world data. This will help commercial banks to avoid monetary losses and prevent financial crisis.

## 1. INTRODUCTION

Credit loans constitute a cornerstone of the banking industry. Stability and profitability of banks rely heavily on the performance of its credit department. This makes detailed analysis of customer's financial background and history an important criterion before making any credit decision. This is a key determinant in reducing the risk involved with credit loans.

Banking management biggest challenge is to evaluate its customers credit risk. In fact, risk estimate is a major factor contributing to any credit decision, and the inability to precisely determine risk adversely affects credit management. This affects both approved and unapproved financing decisions. When credit manager approves a loan, they risk their funds for if customers cannot repay their loan. But if the loan is rejected to a potential customer who could have paid back the loan, then there is a loss and risk of opportunity cost. Wrong decisions result in huge losses for commercial banks which emphasizes that credit risk decisions are key determinants for loan decision making. Due to the significance of credit risk, a number of studies have proposed embracing data mining tools in banks to improve their risk assessment models and hence increase the prediction accuracy of existing models. Data driven machine learning models like genetic algorithms, support vector machines, decision trees and other hybrid models have shown promising results in terms of performance efficiency and accuracy.

As a matter of fact, loan application evaluation at the UAE commercial banks is subjective to nature. This means that every credit risk assessment is done manually thus imposing risk of biasness on personal insights, intuition and knowledge of the credit manager. Even though many banks worldwide have switched on to credit risk analysis, some of them still sticking on to the traditional method and so may face financial crisis or distress. Commercial banks storing their customers data in huge data warehouses can leverage from this with the help of data analytic and mining tools. Despite the increase in the number of non-performing loans and competition in the banking sector, many commercial banks are unwilling to use data driven models to improvise their decision making. Hence, commercial banks in UAE in need to develop more effective machine learning models to improve the performance of its classification accuracy of credit risk assessments. The objective of this research work is to implement data driven machine learning models by applying parameter reduction techniques to classify good creditors from the bad ones and then compare the relative effectiveness of the top models in terms of performance accuracy. In addition, the research focuses on improving credit decision effectiveness and control loan officer tasks as well as save analysis time and cost

## 2. LITERATURE REVIEW

One of the technical factors impacting credit risk evaluation is credit rating (Khashman, 2010). According to (Ghodselahi & Amirmadhi, 2011) credit rating has two kinds of applicants, which are good credit and bad credit. Multilayer feed forward networks are a class of universal approximation (Hornik, Stinchcombe & White, 1989). The machine learning models tend to have a high predictive power (Steiner, Neto, Soma, Shimizu & Nievola, 2006). In fact, data driven networks are basically non parameterized models having a specified degree of precision with one important feature of pattern recognition from highly sophisticated data (Hall, Muljawan, Suprayogi & Moorena, 2009). Pacelli and Azzollini (2011) illustrated that ANN in combination with linear methods have further supported.

Efficiency of neural network model and logistic regression was investigated by Salehi and Mansoury (2011) for forecasting of customer credit risk. The research concluded with both the models having almost the same efficiency. In addition to this, Ghodselahi and Amirmadhi (2011) also implemented and investigated many other data driven hybrid machine learning models in search for their impact on credit rating. They use Support Vector Machine, Neural Networks and Decision Tree as base classifiers. They found out that accuracy of this hybrid model is more than other credit rating methods. Credit rating was also investigated with three methods including Logistic Regression, Neural Networks and Genetic Algorithms (Gouvêa & Gonçalves, 2007). As per this result the performance and efficiency of neural networks and logistic regression are almost similar to each other with neural network being slightly better and genetic algorithms taking the third place. The important roles of ANN in financial application are pattern recognition, classification and time series forecasting (Eletter & Yaseen, 2010).

## 3. RELATED WORK

Author of [16] has built an efficient prediction model for predicting the credibility of customers applying for bank loans. Prototype model using Decision Tree is used to sanction loans to bank customers by predicting the attributes relevant for credibility. The model proposed in [17] has been built using data from commercial banking system in order to validate status of the loans. Out of the three models used namely j48, Bayes Net and naive Bayes, the best accuracy was obtained by using j48. Multi-dimensional is an improved risk prediction clustering algorithm used for determining bad loan applicants as seen in [18]. The research revolved around primary and secondary levels of risk assessments Association Rule in order to avoid redundancy.

Feature selection genetic algorithm was used for a decision tree model in [19]. In addition, the work in [20] proposes two credit scoring models using data mining techniques to support loan decisions for the Jordanian commercial banks. Several non-parametric credit scoring models were involved in working of [21] based on multilayer perceptron approach. The research compares the overall performance of these models with other models build using traditional linear discriminant analysis, logistic regression and quadratic discriminant analysis techniques.

Support vector machine-based credit scoring models using Broad and Narrow default definitions were also built as in [22]. The work showed that data driven models built using Broad default definitions can outperform Narrow default definitions. A brief case study of different data mining techniques like Bayes Classification, bagging algorithm, Random forest, Decision tree, Random Forest and other techniques used in financial data analysis were applied in [23]. The work in [24] checks the applicability of the integrated model on a sample dataset taken from Banks in India. The model was a combination based on the techniques of Logistic Regression, Radial Basis Neural Network, Multi-layer perceptron Model and Support vector machines used in comparing the effectiveness of credit scoring techniques.

Fuzzy Expert System for labelling of credit customers was demonstrated in [25] using Clementine Software. A novel credit scoring model is proposed in 16 that gets an aggregation of classifiers. UCI Machine Learning Repository was used for testing the credit databases in vertical bagging decision tree model with promising results based on accuracy. The research in [26] exploits the predicted behaviour of five classifiers in terms of credit risk prediction accuracy and how such accuracy could be improved.

## 4. DATA COLLECTION AND VARIABLE DEFINATION

Dataset used is the pooled collection of both accepted and rejected applications from different commercial banks of UAE for the year 2016 – 2018 having a debt ratio between 0 – 1. The data content is composed of 7778 cases. In the provided sample, 292 (59.3%) applications were creditworthy while 200 (40.3%) applications were not. The data collection resulted in a total of 11 variables: with 10 being input variables and being the output variable. The definition, coding, type, and descriptive of each of these variables were shown in Table 1.

| Variable | Key | Type | Variable Definition |
|---|---|---|---|
| Resolving the utilization of unsecured lines | P1 | Scale | This attribute indicates the credit card limits of the borrower after excluding any current loan debt and real estate. |
| Age | P2 | Scale | Applicant's age |
| Number of Time30-59 Days Past Due Not Worse | P3 | Scale | The number of this attribute indicates the number of times borrowers have paid their EMIs late but have paid them 30 days after the due date or 59 days before the due date. |
| Debt ratio | P4 | Scale (0 – 1) | If my monthly debt is \$200 and my other expenditure is \$500, then I spend \$700 monthly. If my monthly income is \$1,000, then the value of the debt ratio is \$700/\$1,000 = 0.7000 |
| Monthly income | P5 | Scale | Total monthly income, and used log for transforming it. |
| Number of Open Credit Lines and Loans | P6 | Scale | This attribute indicates the number of open loans and/or the number of credit cards the borrower holds. |
| Number of Times 90 Days Late | P7 | Scale | This attribute indicates how many times a borrower has paid their dues 90 days after the due date of their EMIs. |
| Number Real Estate Loans or Lines | P8 | Scale | This attribute indicates the number of loans the borrower holds for their real estate or the number of home loans a borrower has. |
| Number of Time 60-89 Days Past Due Not Worse | P9 | Scale | This attribute indicates how many times borrowers have paid their EMIs late but paid them 60 days after their due date or 89 days before their due date. |
| Number of Dependents | P10 | Scale | This attribute is self-explanatory as well. It indicates the number of dependent family members the borrowers have. The dependent count is excluding the borrower. |
| Outcome | Output | Binary | Good Creditor or Bad Creditor |

**Table 1:** Dataset variables and its definitions

## 5. METHODOLOGY

With the end of the financial crisis, regulators in the banking industry have put a great focus on risk management and expect financial institutions to have transparent, auditable risk measurement frameworks, depending on portfolio characteristics for regulatory, financial, or business decision-making purposes [27]. Machine learning tools are being developed in order to decode the ever-growing data. Data-driven techniques are being developed to get more insights from data, reduce cost, and increase overall profitability.

Machine learning evolves the study of pattern recognition and computational learning theory in artificial intelligence, which can learn from and make predictions on data. The process starts with an advanced data pre-processing step, where the dataset is cleaned, processed, and unknown values are being averaged out and filled to improve data quality. The process includes data imputation, when relevant, data filtering for very sparse data, and detection and management of outliers. Dataset is then normalized using the Standard Scaler function to reduce and eliminate data reluctance and to improve data integrity. In the next six subsections, we firstly introduce LDA (Linear Discriminant Analysis) which is used as dimensionality reduction techniques; then show four popular and representative ML methods, i.e., 1, 2, 3, 4. We are building an ML model that can help us in order to get an idea, whether a person will be doing any default activity for his loan in the next 2 years.

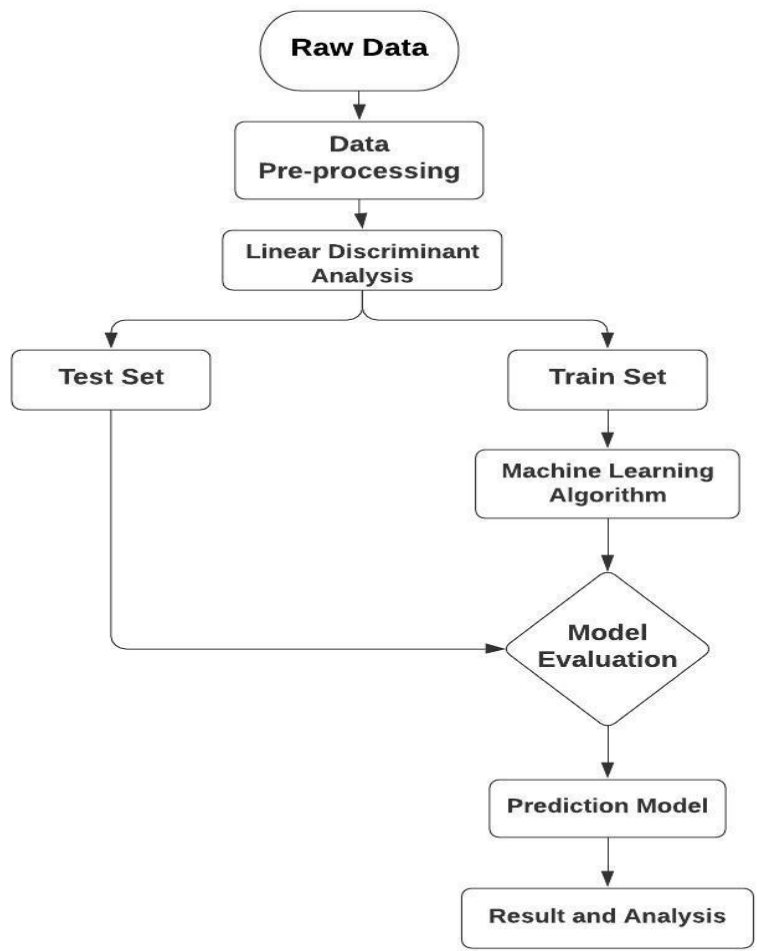

**Figure 1**: Block Diagram of the proposed model

## 6. DATA PREPROCESSING

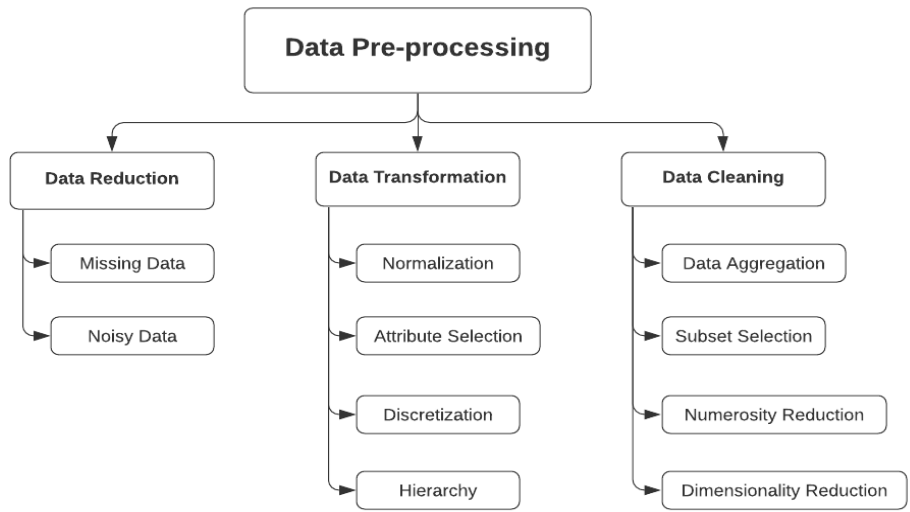

**Figure 2:** Data pre-processing

**6.2 Data Cleaning**
Data Cleaning plays an important role in the field of Analytics and Machine Learning [28]. Data Cleaning means the process of identifying the incorrect, incomplete, inaccurate, irrelevant or missing part of the data and then modifying, replacing or deleting them according to the necessity. If the data is corrupted then it may hinder the process or provide inaccurate results.

**6.3 Data Transformation**
Data transformation is the process in which you take data from its raw, siloed and normalized source state and transform it into data that's joined together, dimensionally modelled, de-normalized, and ready for analysis. Data transformation can be time-consuming, expensive, and tedious [29]. Nevertheless, transforming your data will ensure maximum data quality which is imperative to gaining accurate analysis, leading to valuable insights that will eventually empower data-driven decisions.

**6.4 Data Reduction Using Linear Discriminant Analysis (LDA)**
LDA, is a linear machine learning algorithm used for multi-class classification. Linear Discriminant Analysis seeks to best separate (or discriminate) the samples in the training dataset by their class value [30]. Specifically, the model seeks to find a linear combination of input variables that achieves the maximum separation for samples between classes (class centroids or means) and the minimum separation of samples within each class. We can use LDA to calculate a projection of a dataset and select a number of dimensions or components of the projection to use as input to a model.

Linear discriminant analysis is used as a tool for classification, dimension reduction, and data visualization. It has been around for quite some time now. Despite its simplicity, LDA often produces robust, decent, and interpretable classification results. A classifier with a linear decision boundary, generated by fitting class conditional densities to the data and using Bayes' rule [31] [32]. The model fits a Gaussian density to each class, assuming that all classes share the same covariance matrix. The fitted model can also be used to reduce the dimensionality of the input by projecting it to the most discriminative directions, using the transform method.

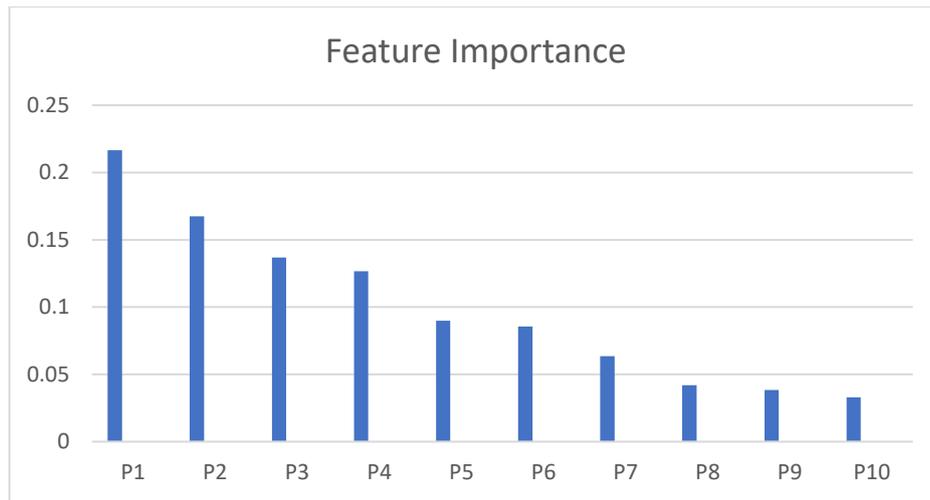

**Figure 3**: Correlation of the attributes with the output

After the implementation of LDA on the dataset and the extraction of the features as shown in figure 3, attributes with decreasing order of correlation with the output are Resolving utilization of unsecured lines, age, NumberOfTime30-59DaysPastDueNotWorse, debt ratio, monthly income, Number of Open Credit Lines and Loans, NumberOfTimes90DaysLate, Number Real Estate Loans or Lines, NumberOfTime60-89DaysPastDueNotWorse, and Number of Dependents respectively. The scikit-learn library provides the Linear Discriminant Analysis class that can be fit on a dataset and used to transform a training dataset and any additional dataset in the future.

## 7. MACHINE LEARNING MODEL

### 7.1 Logistic Regression
Logistic regression, despite its name, is a linear model for classification rather than regression. Logistic regression is also known in the literature as logit regression, maximum-entropy classification (MaxEnt), or the log-linear classifier. In this model, the probabilities describing the possible outcomes of a single trial are modelled using a logistic function. Logistic regression is implemented in Logistic Regression. This implementation can fit the binary, One-vs-Rest, or multinomial logistic regression with optional $\ell 1$, $\ell 2$ or Elastic-Net regularization.

### 7.2 AdaBoost Classifier
An AdaBoost classifier is a meta-estimator that begins by fitting a classifier on the original dataset and then fits additional copies of the classifier on the same dataset but where the weights of incorrectly classified instances are adjusted such that subsequent classifiers focus more on difficult cases. This class implements the algorithm known as AdaBoost-SAMME

### 7.3 Decision Tree
Decision Trees (DTs) are a non-parametric supervised learning method used for classification and regression. A tree can be seen as a piecewise constant approximation. It builds the model in a stage-wise fashion like other boosting methods do, and it generalizes them by allowing optimization of an arbitrary differentiable loss function.

### 7.4 Neural Network
A neural network is a class of feedforward artificial neural network (ANN). The term MLP is used ambiguously, sometimes loosely to *any* feedforward ANN, sometimes strictly to refer to networks composed of multiple layers of perceptron. An MLP consists of at least three layers of nodes: an input layer, a hidden layer and an output layer. Except for the input nodes, each node is a neuron that uses a nonlinear activation function. MLP utilizes a supervised learning technique called backpropagation for training.[2][3] Its multiple layers and non-linear activation distinguish MLP from a linear perceptron. It can distinguish data that is not linearly separable.

## 8. EXPERIMENTATION AND RESULTS

All the four selected models - Logistic Regression, Adaboost Classifier, Decision Trees and Neural Network – were trained on the dataset containing 7778 customer records of commercial banks involving credit risk assessment decisions. These models were trained on the dataset containing attributes with the highest correlation with the final binary output. This was further divided into three cases which revolved around the number of parameters used as shown in Table 2

| Number of Attributes used | Attributes involved |
|---|---|
| 2 | Resolving utilization of unsecured lines age |
| 3 | Resolving utilization of unsecured lines age NumberOfTime3059DaysPastDueNotWorse |
| 4 | Resolving utilization of unsecured lines age NumberOfTime3059DaysPastDueNotWorse debt ratio |

**Table 2:** Attributes involved in model evaluation

Thus, when number of attributes used were 2, parameters used were 'Resolving utilization of unsecured lines (P1)' and 'age (P2)'; when number of attributes used were 3, parameters used were 'Resolving utilization of unsecured lines (P1)', 'age (P2)' and 'Number of time 30-59 Days past due not worse (P3)'; when number of attributes used were 4, parameters used were 'Resolving utilization of unsecured lines (P1)', 'age (P2)', and 'debt ratio (P3)' respectively. After initializing the attributes for the testing and training set, the model was trained on the four Machine Learning models: Logistic Regression, Adaboost Classifier, Decision Trees and Neural Network. These models were compared on the basis of their accuracy on the testing set, both before and after implementing Linear Discriminant Analysis on the dataset.

| Model | Accuracy Before LDA (%) | | | Accuracy After LDA (%) | | |
|---|---|---|---|---|---|---|
| | Number of Parameter | | | Number of Parameters | | |
| | 2 | 3 | 4 | 2 | 3 | 4 |
| Logistic Regression | **93.739** | 92.115 | 92.522 | **95.026** | **95.066** | 95.144 |
| Adaboost Classifier | 92.204 | **92.638** | 93.154 | 95.009 | 95.023 | **95.200** |
| Decision Trees | 92.488 | 92.455 | 93.451 | 94.921 | 94.918 | 95.042 |
| Neural Network | 93.291 | 92.007 | **93.633** | 94.985 | 94.999 | 95.186 |

**Table 3:** Models accuracy before and after LDA

After model evaluation and testing, accuracy obtained in each case was noted down as shown in Table 3. Logistic Regression model had the highest accuracy of 93.739% as compared to other three models before the implementation of LDA. Also, Logistic Regression obtained the highest accuracy when number of parameters used were 2 (P1, P2). Adaboost Classifier and neural network had the highest accuracy of 92.638% and 93.633%, when number of parameters used were 3 (P1, P2, P3) and 4 (P1, P2, P3, P4) respectively.

After the implementation of LDA, Adaboost Classifier showed the highest accuracy of 95.200% as compared to other models. Number of parameters used in this case were 4 (P1, P2, P3, P4). In addition, Logistic Regression showed the highest accuracy of 95.026% and 95.066% when number of parameters used were 2 (P1, P2) and 3 (P1, P2, P3) respectively. Further, in case of LDA, highest accuracy is obtained when the number of parameters used are 4 – P1, P2, P3, P4 – whereas the highest accuracy is obtained when the number of parameters used are 2 – P1, P2 – without the application LDA.

## CONCLUSION AND FURTHER RESEARCH

In this paper, we focus on increasing the accuracy of credit risk assessment data-driven model using LDA reduction technique for commercial banks of UAE. Accurate credit risk prediction model can help bank prevent huge losses and avoid financial crisis. A lot of commercial banks have their credit risk model constructed on top traditional statistical methods. Even though traditional models are quite robust and efficient, they are significantly behind as compared to current machine learning models when it comes performance accuracy.

All the four models – Logistic Regression, Adaboost Classifier, Decision Tree and Neural Network - have shown promising results with very negligible difference between their accuracy performance. However, there is a significant difference in accuracy before and after applying LDA on the dataset. This indicates that applying LDA on the dataset increases the performance of the machine learning models in terms of accuracy. Thus, bank policy makers and credit risk decisions should introduce LDA as a dimensionality reduction tool to identify bad creditors from good creditors. Further, parameters namely Resolving the utilization of unsecured lines, age, Number of Time30-59 Days Past Due Not Worse and debt ratio were found to have the most correlation with the outcome of credit risk assessment model indicating significant importance must be given to these four parameters when making a credit risk decision in order to avoid the possibility of financial distress and business failure.

## ACKNOWLEDGEMENT

This work was supported by Birla Institute of Technology and Science, Pilani, Dubai Campus, Dubai, UAE. Special thanks to Professor Dr Parizad Dungore, Assistant Professor, Finance and Accountancy, Department of Humanities and Social Science, BITS Pilani Dubai Campus.